\documentclass{aa}
\usepackage{graphics}
\usepackage{times}
\def\kms{km\,s$^{-1}$}
\def\Am{\AA\,mm$^{-1}$}
\def\hb{H$\beta$}
\def\ha{H$\alpha$}
\def\nii{\lbrack\ion{N}{ii}\rbrack}
\def\oiii{\lbrack\ion{O}{iii}\rbrack}
\def\feii{\ion{Fe}{ii}}

\def\roiii{$\lambda$5007$/$H$\beta$}

\def\rnii{$\lambda$6583$/$H$\alpha$}

\def\hii{\ion{H}{ii}}

\begin{document}

\thesaurus{11                
	      (11.01.2;      
	       11.19.1;      
	       11.09.1)}     
%
\title{The emission-line spectrum of KUG 1031+398 and the 
Intermediate Line Region
\thanks{Based on observations collected at the Observatoire de Haute-Provence
(CNRS), France.}
}
\author{A.\,C. Gon\c{c}alves \and P. V\'eron \and M.-P. V\'eron-Cetty}
\offprints{A.\,C. Gon\c{c}alves, {\small{anabela@obs-hp.fr}} }
\institute{Observatoire de Haute-Provence (CNRS), F-04870 Saint Michel
l'Observatoire, France}
\date{Received~~/~~Accepted}
\maketitle

\begin{abstract}
We present results based on the analysis of optical spectra of 
\object{KUG 1031+398}, 
a Narrow Line Seyfert 1 (NLS1) galaxy for which Mason et al. (1996) reported
evidence for a line-emitting region ``intermediate'' (both in terms of
velocity and density) between the conventional broad and narrow line regions
(BLR and NLR, respectively). From our observations and modelling of the
spectra, we get a consistent decomposition of the line profiles into four
components: an extended \hii\ region with unresolved lines, two distinct
Seyfert-type clouds identified with the NLR, and a relatively narrow
``broad line'' component emitting only Balmer lines but no forbidden lines.
Therefore, and although we find this object to be exceptional in having
line-emission from the BLR with almost the same width as the narrow lines,
our interpretation of the data does not support the existence of an
``intermediate'' line region (ILR).

\keywords{galaxies: active -- galaxies: Seyfert -- galaxies: individual: 
KUG~1031+398} 
\end{abstract}

\section{Introduction}
\subsection{The Intermediate Line Region}
It is commonly accepted that line-emission in AGNs comes from two well
separated regions: one, compact (smaller than 1 pc) and lying close to 
the central engine, has a high electron density ($N_{\rm e}$ $>$ 10$^{8}$ 
cm$^{-3}$) and is responsible for the 
production of broad (FWHM $\sim$ thousands of \kms) permitted lines -- 
the BLR; the other, more extended and lying further 
away from the central source (10--1\,000 pc), has lower electron densities 
(10$^{3} \leq N_{\rm e} \leq$ 10$^{5}$ cm$^{-3}$) 
and emits lines with a lower velocity dispersion ($\sim$ hundreds of \kms) 
-- the NLR. 

A line-emission ``gap'' is usually observed between the two regions, 
most objects showing an optical spectrum which can be fitted by line 
profiles corresponding to
clouds belonging to one or the other line-emitting regions. This line-emission
gap can be explained by the presence of dust mixed with the gas (Netzer \&
Laor 1993). Nevertheless, the existence of an intermediate region, both in
terms of velocity and density, is expected; in such a region, the \oiii\ lines 
would be partially collisionaly de-excited\footnotemark[1] and show 
substantially broadened wings (Shields 1978). This ILR should not be confused 
with the ILR found in QSOs by Brotherton et al. (1994), which is much smaller 
and denser, with a velocity dispersion of the order of 2\,000 \kms\ and 
density $\sim$ 10$^{10}$ cm$^{-3}$.
\footnotetext[1]{If we make the assumptions that the 
excitation conditions in both the NLR and the ILR are the same and that in 
the ILR, \roiii\ $\sim$ 1, the \oiii$\lambda$5007 line is collisionaly 
de-excited by about a factor 10. According to the formula given by 
Seaton (1975), this implies a density of 1--3\,10$^{6}$ cm$^{-3}$ if the 
electronic temperature is in the range 1--3\,10$^{4}$ K.}

Mason et al. (1996) presented high-resolution (2 \AA) 
optical spectroscopic observations of KUG 1031+398. The model they used to 
fit the data revealed a line-emitting region with lines of intermediate width 
(FWHM $\sim$ 1\,000 \kms); according to Mason et al., this region would 
dominate the Balmer lines profile, being also a significant contributor to 
the \oiii $\lambda\lambda$4959, 5007 lines, with a flux ratio 
\roiii\ = 1.4, suggesting an intermediate density. 

Osterbrock (1978) thought that he had detected, in a few Seyfert 1 galaxies, 
faint wings to the \oiii\ lines with essentially the same widths as the 
Balmer lines. Crenshaw \& Peterson (1986) and van Groningen \& de Bruyn 
(1989) have found broad wings in the \oiii\ lines of a number of Seyfert~1 
galaxies, implying the presence, in these objects, of an ILR with a density 
of a few times 10$^{6}$ cm$^{-3}$, similar to the one reported in 
KUG 1031+398; however, all these objects show strong \feii\ emis\-sion, and 
the observed broad \oiii\ components could be due to an inaccurate removal 
of the \feii\ blends (Boroson \& Green 1992).

In summary, although the presence in Seyfert~1 galaxies of emitting clouds 
with density intermediate between those of the ``broad'' and ``narrow''
components is not unexpected, no uncontroversial report of the existence of
such intermediate components has ever been made to the best of our
knowledge. Therefore, the claims by Mason et al. (1996) that the NLS1 
KUG 1031+398 shows evidence for an ILR induced us to conduct new 
spectroscopic observations and modelling of its emis\-sion-line features. 

\subsection{KUG 1031+398}
2RE J1034+396 was found in the {\it ROSAT} Wide Field Camera all-sky
extreme-ultraviolet survey (Pounds et al. 1993; Pye et al. 1995).
It was identified by Shara et al. (1993) and Mason et al. (1995)
with the compact UV-excess 15.0 mag galaxy KUG 1031+398 (Takase \&
Miyauchi-Isobe 1987) at $z$ = 0.042. This object has an intense soft
X-ray emission with an unusually steep 2--10 keV power law of photon index
$\Gamma \sim$ 2.6 $\pm$ 0.1 (Pounds et al. 1995) and an even steeper
0.1--2.4 keV power law with $\Gamma$ = 3.4 $\pm$ 0.3 (Puchnarewicz et al.
1995) or $\Gamma$ = 4.4 $\pm$ 0.1 (Rodriguez-Pascual et al. 1997). 

UV spectroscopy with the {\it Hubble Space Telescope} shows the 
Ly\,$\alpha$ profile to be complex, with a narrow (400 \kms\ FWHM) and 
a broad (1\,600 \kms\ FWHM) component (Puch\-narewicz et al. 1998). The 
optical continuum is not polarized (Breeveld \& Puchnare\-wicz 1998). 

The broad component of the Balmer lines is relatively narrow (FWHM $\sim$
1\,500 \kms) and, consequently, this object has been classified as a NLS1
by Puchnarewicz et al. (1995). Narrow line Seyfert 1 galaxies are defined 
as Seyfert~1s having ``broad'' Balmer lines narrower than 
2\,000 \kms\ FWHM (Osterbrock 1987). Most NLS1s have a steep 
soft ($<$ 1 keV) X-ray component and, conversely, 
most ultra-soft X-ray sources are associated with a NLS1 (Puch\-narewicz et 
al. 1992; Greiner et al. 1996; Boller et al. 1996; Wang et al. 1996). 

\section{Observations}
Spectroscopic observations of KUG 1031+398 were carried out with the
spectrograph CARELEC (Lema\^{\i}tre et al. 1989) attached to the
Cassegrain focus of the Observatoire de Haute Provence (OHP) 1.93 m telescope.
The detector was a 512$\times$512 pixels, 27$\times$27 $\mu$m Tektronix CCD.
We used a 600 l\,mm$^{-1}$ grating giving a dispersion of 66 \Am.
On January 10, 1997 we obtained a 20 min exposure in the range
$\lambda\lambda$\,6175--7075 \AA, on March 4, a 20 min exposure in the
range $\lambda\lambda$\,4780--5780 \AA, and three more on March 5.

The slit width was 2\farcs1, corresponding to a projected slit width on the
detector of 52 $\mu$m, or 1.9 pixel. The resolution, as measured on the
night sky emission lines, was 3.4 \AA\ FWHM in the blue and 3.5 \AA\ in the
red. In both cases the galaxy nucleus was centered on the slit and 3 columns 
of the CCD ($\sim$ 3\farcs2) were extracted, corresponding to $\sim$ 4 kpc 
at the distance of the galaxy (with $H\rm _{o}$ = 50 \kms\,Mpc$^{-1}$). 

\begin{figure*}
\resizebox{13.1cm}{15cm}{\includegraphics{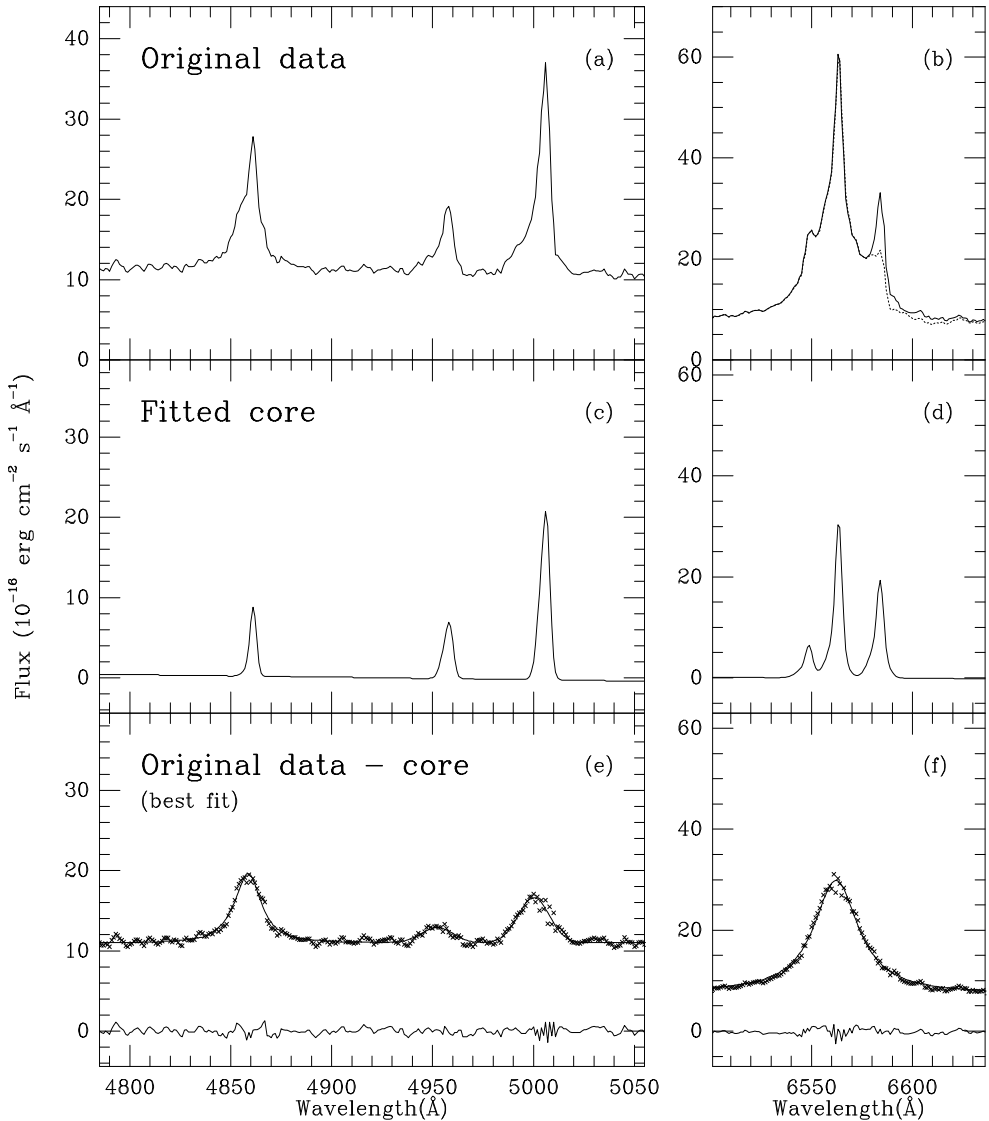}}
\hfill
\parbox[b]{45mm}{ 
\caption{\label{Obs_and_Fits} 
Blue (a) and red (b) spectra of KUG 1031+398 in the rest frame; 
in (b) we also give the spectrum before correcting for the atmospheric 
absorption (dotted line). The narrow line-core components (c and d) were fitted 
with Gaussians and subtracted from the original data, the result being shown 
in (e) to (h). In (e) and (f), we show our best fit (solid line) together 
with the data points (crosses); the lower solid lines represent the residuals.}
}
\end{figure*}

The spectra were flux calibrated using the standard stars EG 247 (Oke
1974) and Feige 66 (Massey et al. 1988), observed with the same instrumental 
settings; these standards were also used to
correct the red spectrum for the atmospheric B band at $\lambda$6867 \AA\
(Fig. \ref{Obs_and_Fits}b). 

\section{Analysis}
\subsection{Methodology}
Positive correlations between line-widths and ionization
potentials/critical densities have been observed in the narrow line 
region of many Seyfert galaxies. Negative correlations are also sometimes 
found. A positive correlation implies that the density and/or ionization 
parameter gradually increases inward in the NLR of these objects. In the case 
of correlations with critical densities, the observed values range from 
$\sim$ 10$^{3}$ to $\sim$ 3\,10$^{6}$ cm$^{-3}$ (Filippenko \& Sargent 1985).

Sub-structures were found in the narrow line profiles of most objects 
suggesting that the line emitting region is a collection of individual 
clouds in motion relative to each other and producing different parts of the 
line profiles (see for instance Veilleux 1991, Espey et al. 1994 or 
Ferguson et al. 1997). 

These findings induced us to assume that each 
of these clouds is characterized by a single density and that lines 
coming from the same emission-region should have the same profile 
and mean velocity. We therefore tried to model the spectra of 
KUG 1031+398 with the smallest possible
number of line sets, each set including three Gaussians (modelling \ha\ and
the \nii $\lambda\lambda$6548, 6583 lines, or \hb\ and the
\oiii $\lambda\lambda$4959, 5007 lines) having the same velocity shift
and width, with the additional constraint that the intensity ratio of the
two \nii\ (respectively \oiii) lines was taken to be equal to the theoretical
value of 3.00 (respectively 2.96) (Osterbrock 1974). In a physically
meaningful and self-consistent model, the components found when fitting the
blue and red spectra should have velocity shifts and widths compatible
within the measurement errors.

\subsection{The narrow line-core components}
The spectra were deredshif\-ted assuming $z$ = 0.0434 
(Fig. \ref{Obs_and_Fits}a and b) and 
analysed in terms of Gaussian components as described above. We discovered 
first that the core of the lines could not be fitted by a single set of 
narrow Gaussian profiles. To get a satisfactory fit, two sets of Gaussian 
components are needed: the first, unresolved (and subsequently taken as the 
origin of the velocity scales) has \rnii\ = 0.55 and \roiii\ = 1.27 and 
corresponds to a \hii\ region (Fig. \ref{CCD_Hbeta}); the second is resolved 
(FWHM $\sim$ 350 \kms, 
corrected for the instrumental broadening), blueshif\-ted by $\sim$ 95  
\kms\ with respect to the narrower components and has line intensity ratios 
typical of a Seyfert 2 (\rnii\ = 0.84, \roiii\ = 10.2). 

\subsection{The ``broad'' components}
At this stage, we removed from the blue and red spectra the best fitting 
line-core (the \hii\ region and the Seyfert 2 nebulosity, 
Fig. \ref{Obs_and_Fits}c and d), 
obtaining two spectra we shall call ``original data minus core''. The blue one 
was then fitted 
with a broad \hb\ Gaussian component and two sets of three components 
modelling the narrow \hb\ and \oiii $\lambda\lambda$4959, 5007 lines. The 
result is very suggestive: one set has a strong \hb\ line and very weak 
negative \oiii\ components, while the other set displays a strong \oiii\ 
contribution and a weak negative \hb\ component, showing that we have 
in fact a \hb\ component with no associated \oiii\ emission and \oiii\ 
lines with a very weak (undetected) associated \hb; in other words, the 
region producing the \hb\ line does not emit forbidden lines, while the 
\oiii\ emitting region has a high \roiii\ ratio, which are, respectively, 
the characteristics of the ``broad'' and ``narrow'' line regions in 
Seyfert~1 galaxies. 

Having these results in mind, we optimized this last fit by using a Lorentzian 
profile for the \hb\ line, with no associated \oiii\ emission, and a set of 
three Gaussians for the remaining contribution coming from the ``narrow'' 
components (this is not the first time Lorentzian profiles are used to fit AGN 
emission lines; for example St\"uwe et al. (1992) found that, in the case of 
NGC 4258, the narrow lines were better fitted by Lorentzians, rather 
than Gaussians). 

The best fit is presented in Fig. \ref{Obs_and_Fits}e: in this model, 
the flux of the ``narrow'' (Gaussian) \hb\ component is only 
9\% of the ``broad'' (Lorentzian) \hb\ component. 
The \hb\ Lorentzian component is blueshif\-ted by 150 \kms\ with a wi\-dth of 
915 \kms. The Gaussian components are 
blueshif\-ted by $\sim$ 395 \kms\ and their width is $\sim$ 1\,115 
\kms\footnotemark. Mason et al. found a FWHM = 1\,030 $\pm$ 150 
\kms\ for this component which is blueshif\-ted by 240 $\pm$ 30 \kms. This 
blueshift, however, is measured with respect to the \oiii\ lines core which 
is dominated by the Seyfert 2 cloud, itself blueshif\-ted by 95 \kms\ with 
respect to the \hii\ region; the blueshift of Mason et al.'s intermediate 
component is, therefore, 240 + 95 = 335 \kms, in agreement with our value 
of 395 \kms.

In Seyfert 1 galaxies, the Balmer decrement of the broad component is never 
smaller than that of the narrow component; in the present 
case, we therefore expect the ``narrow'' \ha\ component flux 
to be less 
than 9\% of the ``broad'' \ha\ component flux. Moreover, the 
\nii$\lambda$6583 line flux is, in Seyfert galaxies, equal or smaller than the 
narrow \ha\ component flux. So, in KUG 1031+398, we expect the narrow lines 
to be quite weak compared to the ``broad'' \ha\ component, and we fitted the 
``original data minus core'' red spectrum with a single Loren\-tzian profile 
of 1\,205 \kms\ FWHM, blueshifted by 65 \kms\ with respect to the 
\hii\ region. This fit is shown in Fig. \ref{Obs_and_Fits}f. 

Another model allowing, in addition, for a set of \ha\ and \nii\ Gaussian 
components was also tested, resulting in a fit of similar quality; the 
very small \rnii\ ratio observed for this solution (0.2), shows that 
the nitrogen lines may be considered as undetectable.

\section{Results and Discussion}
\begin{figure}[t]
\resizebox{3.5cm}{!}{\includegraphics{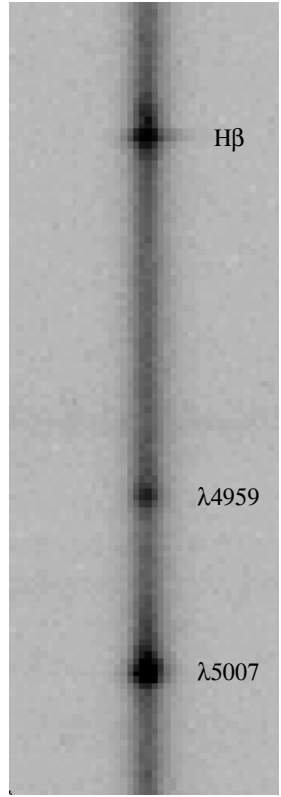}}
\hfill
\parbox[b]{50mm}{
\caption{\label{CCD_Hbeta} 
Enlargement of part of the CCD frame showing the \hb\ and 
\oiii$\lambda\lambda$4959, 5007 lines in the spectrum of KUG 1031+398 
(average of three 20 min exposures, after cosmic rays removal); the 
extended (narrow) component of \hb\ is clearly visible.}
}
\end{figure}

Our new observations and modelling of KUG 1031+398 yield a consistent
decomposition of the emission-line profile into four components (see
Table \ref{results}):

\begin{enumerate}
\item An extended \hii\ region with unresolved lines (Fig. \ref{CCD_Hbeta});
\item A first Seyfert-type cloud with relatively narrow lines 
($\sim$ 350 \kms\ FWHM), blueshif\-ted by 95 \kms, belonging to the NLR; 
\item A second Seyfert-type component with somewhat broader lines, 
blueshif\-ted by $\sim$ 395 \kms; the width of the lines in this component 
($\sim$ 1\,115 \kms\ FWHM), which may seem large for a Seyfert 2, is not 
exceptional as the FWHM of the lines in the prototype Seyfert 2 galaxy NGC 1068 
is $\sim$ 1\,670 \kms\ (Alloin et al. 1983). Only the \oiii\ lines are 
observed in this component, with \roiii\ $\sim$ 6.1, a line ratio 
characteristic of NLRs; 
\item Finally, a Narrow Line Seyfert~1 cloud with lines well fitted
by a Lorentzian profile of $\sim$ 1\,060 \kms\ FWHM, blueshif\-ted by 105 \kms.
\end{enumerate}

Our analysis shows that the emission line spectrum of KUG 1031+398 
can be satisfactorily decomposed in a set of components which have 
line ratios characteristics of \hii\ regions or conventional NLR or 
BLR clouds, whithout the need to invoke the presence of an ILR 
characterized by \roiii\ $\sim$ 1.

There are two main reasons why our analysis yields different results from
those published by Mason et al. (1996). First, KUG 1031+398 having a 
redshift of $\sim$ 0.043, the \nii $\lambda$ 6583 line coincides with the 
atmospheric B band. When correcting for this absorption
feature, the \nii\ true intensity is recovered (Fig. \ref{Obs_and_Fits}b) 
and our red spectrum appears different from the published one; different 
line-ratios and widths are therefore not unexpected.

Second, the line-profile analysis of Mason et al. differs from ours in that,
while we force each Balmer component to be associated with forbidden lines
having the same velocity and width, Mason et al. allow these parameters to
have different values for the Balmer and forbidden line components. As a
result, they found three \hb\ components (a narrow, an intermediate and a
broad one), as well as two \oiii\ components (a narrow and an intermediate
one); they also detected three \ha\ components (again a narrow, an
intermediate and a broad one), but only a single \nii\ component (narrow).
The measured width of the narrow \hb\ component is 150 $\pm$ 20 \kms\
FWHM, while the width of the narrow \oiii\ lines is 265 $\pm$ 10 \kms; this
last value, significantly larger than the narrow \hb\ line width, suggests
that the \oiii\ lines may have a complex profile. Moreover, the width of
the \nii\ lines is found to be significantly larger 
(400 $\pm$ 60 \kms) than that of the narrow \ha\ component
(190 $\pm$ 40 \kms); this could be due to an inaccurate correction of
the atmospheric B band, as we have seen above.

Although our spectra have a lower resolution than those obtained by Mason et
al. (3.4 \AA\ compared to 2 \AA\ FWHM), this does not affect the analysis;
the narrow core components being identified and subtracted, all the discussion
is centered on the broader components, well resolved even with our lower
resolution. Similarly, the larger slit width used in our observations
(2\farcs1 compared to 1\farcs5 for Mason et al.) does not affect the study
of these broader components, since only the contribution from the extended
emitting region (the \hii\ region, Fig. \ref{CCD_Hbeta}), removed with the 
core, changes with the slit width.

Boller et al. (1996) and Wang et al. (1996) have suggested 
that the small width of the broad Balmer
lines and the soft X-ray excess characteristic of NLS1 galaxies could be
the effect of a high accretion rate on an abnormally small mass black hole.
Mason et al. (1996) have argued that, although the emission line spectrum in 
KUG 1031+398 is dominated by the ILR, a weak broad component is present with
line-widths of the order of 2\,500 km\,s$^{-1}$ FWHM and that, therefore, at
least in this object, such a model is not required. 

Our analysis of the spectra of KUG 1031+398 has shown that, in the BLR, 
the Bal\-mer lines are well fitted by a Lorentzian profile with $\sim$ 
1\,060 km\,s$^{-1}$ FWHM; this 
value is much narrower than the value found by Mason et al. ($\sim$ 2\,500 
\kms). This is due to the fact that we used a Lorentzian, rather than 
a Gaussian profile to fit the broad Balmer lines; the Lorentzian profile 
was required by the presence of broad wings, fitted with a Gaussian 
by Mason et al. (1996). 

We have shown (Gon\c{c}al\-ves et al. in preparation) that in NLS1s the 
broad component of the Balmer lines is generally better fitted by a 
Lorentzian than by a Gaussian; the Lorentzian Balmer lines (component 4), 
without any measurable 
associated forbidden line, would qualify this object as a NLS1 with, 
in fact, very narrow lines. So, in this respect, KUG 1031+398 is a 
normal NLS1 and could be explained by the same small 
black hole mass model as the other objects of this class. 

\begin{table}[h]
\begin{center}
\caption{\label{results} 
Emission line profile analysis of KUG 1031+398. I(\hb) and 
I(\ha) are in units of 10$^{-16}$ erg\,s$^{-1}$\,cm$^{-2}$. In columns 1 
and 2 we give the mean of the relative velocities and widths measured on the 
blue and red spectra. The FWHMs 
are corrected for the instrumental broadening.}
\begin{flushleft}
\begin{tabular}{p{0.15cm}rrrcrr}
\hline
 & $\Delta$V\verb+  + & FWHM\,    &  \underline{$\lambda$5007\,} 
 & \underline{$\lambda$6583\,}    &  I(\hb) & I(\ha) \\
 & (\kms) & (\kms) & \hb\verb+ +  &  ~\ha &  &  \\
\hline
1       &   0\verb+  +       &   $<$80\verb+ +      &  1.27           & 
0.55    &   29$\:\,$         &   93\verb+ +  \\
2       &   $-$ 95\verb+  +  &   350\verb+ +        &  10.2\verb+ +   & 
0.84    &   8$\:\,$          &   81\verb+ +  \\
3       &   $-$ 395\verb+  + &   1115\verb+ +       &   6.1\verb+ +   & 
--      &   18$\:\,$         &   -- \verb+ + \\
4       &   $-$ 105\verb+  + &   1060\verb+ +       & --\verb+  +     &    
--      &   189$\;$          &   938\verb+ + \\
\hline
\end{tabular}
\end{flushleft}
\end{center}
\end{table}

\section{Conclusions}
We have obtained new spectra of KUG 1031+398 around \hb\ and \ha. We have 
found that the emission-line spectrum of this object can be modelled with 
four components: an extended \hii\ region, two narrow emission 
regions of Seyfert 2-type and a relatively narrow ``broad line'' component, 
well fitted by a Lorentzian profile. 

We disagree with Mason et al. on the analysis of the emission line
profile of KUG 1031+398, in the sense that we find no evidence for the
presence of an ``intermediate'' component in which the forbidden lines are 
almost, but not completely, suppressed by collisional de-excitation. 
Nevertheless, we
find that this object is exceptional in having a ``narrow'' line region
(defined as a region where \roiii\ $\ge$ 5) with almost the same width 
at half maximum as the ``broad'' line region (Bal\-mer lines with no 
detectable associated forbidden lines); however, in the first case, the 
line-profile is Gaussian, while in the second case, it is Lorentzian.


\begin{acknowledgements} 
We would like to thank A. Rodriguez-Ardila and G. Shields for useful comments 
and suggestions. A.\,C. Gon\c{c}alves acknowledges support from the 
{\it Funda\c{c}\~ao para a Ci\^encia e a Tecnologia}, Portugal, during 
the course of this work (PRAXIS XXI/BD/5117 /95 PhD. grant). 
\end{acknowledgements}



\begin{thebibliography}{}





\bibitem{}Alloin D., Pellat D., Boksenberg A., Sargent W.\,L.\,W., 1983,  
ApJ 275, 493













\bibitem{}Boller T., Brandt W.\,N., Fink H., 1996, A\&A 305, 53


\bibitem{}Boroson T.\,A., Green R.\,F., 1992, ApJS 80, 109



\bibitem{}Breeveld A.\,A., Puchnarewicz E.\,M., 1998, MNRAS 295, 568



\bibitem{}Brotherton M.\,S., Wills B.\,J., Francis P.\,J., 
Steidel C.\,J., 1994, ApJ 430, 495









\bibitem{}Crenshaw D.\,M., Peterson B.\,M., 1986, PASP 98, 185













\bibitem{}Espey B.\,R., Turnshek D.\,A., Lee L. et al., 1994, ApJ 434, 484





\bibitem{}Ferguson J.\,W., Korista K.\,T., Baldwin J.\,A., Ferland G.\,J., 
1997, ApJ 487, 122




\bibitem{}Filippenko A.\,V., Sargent W.\,L.\,W., 1985, ApJS 57, 503












\bibitem{}Greiner J., Danner R., Bade N. et al., 1996, A\&A 310, 384















































\bibitem{}Lema\^{\i}tre G., Kohler D., Lacroix D., Meunier J.-P., Vin A., 1989, 
A\&A 228, 546



























\bibitem{}Mason K.\,O., Hassall B.\,J.\,M., Bromage G.\,E. et al., 1995, 
MNRAS 274, 1194

\bibitem{}Mason K.\,O., Puchnarewicz E.\,M., Jones L.\,R., 1996, MNRAS 283, L26

\bibitem{}Massey P., Strobel K., Barnes J.\,V., Anderson E., 1988, ApJ 328, 315






  
\bibitem{}Netzer H., Laor A., 1993, ApJ 404, L51



\bibitem{}Oke J.\,B., 1974, ApJS 27, 21



\bibitem{}Osterbrock D.\,E., 1974, Astrophysics of gaseous nebulae., 
Freeman and company, San Francisco

\bibitem{}Osterbrock D.\,E., 1978, {\it Physica Scripta} 17, 285




\bibitem{}Osterbrock D.\,E., 1987, Lecture Notes in Physics 307, 1















\bibitem{}Pounds K.\,A., Allen D.\,J., Barber C., et al., 1993, MNRAS 260, 77

\bibitem{}Pounds K.\,A., Done C., Osborne J.\,A., 1995, MNRAS 277, L5

\bibitem{}Puchnarewicz E.\,M., Mason K.\,O., Cordova F.\,A. et al., 1992, 
MNRAS 256, 589

\bibitem{}Puchnarewicz E.\,M., Mason K.\,O., Siemiginowska A., Pounds K.\,A., 
1995, MNRAS 276, 20

\bibitem{}Puchnarewicz E.\,M., Mason K.\,O., Siemiginowska A., 1998, MNRAS 
293, L52

\bibitem{}Pye J.\,P., McGale P.\,A., Allan D.\,J. et al., 1995, MNRAS 274, 1165
 





\bibitem{}Rodriguez-Pascual P.\,M., Mas-Hesse J.\,M., Santos-Ll\'eo M., 1997, 
A\&A, 327, 72













\bibitem{}Seaton M.\,J., 1975, MNRAS 170, 475



\bibitem{}Shara M.\,M., Shara D.\,J., McLean B., 1993, PASP 105, 387

\bibitem{}Shields G.\,A., 1978. In: Wolfe A.M. (ed.), Pitsburgh conference 
on BL Lac objects, University of Pittsburgh, Pittsburgh, Pennsylvania, p. 257















\bibitem{}St\"uwe J.\,A., Schulz H., Huehnermann H., 1992, A\&A 261, 382




\bibitem{}Takase B., Miyauchi-Isobe N., 1987, Ann. Tokyo astron. Obs. 
2nd series 21, 363






\bibitem{}van Groningen E., de Bruyn A.\,G., 1989, A\&A 211, 293

\bibitem{}Veilleux S., 1991, ApJ 369, 331



















 

\bibitem{}Wang T., Brinkmann W., Bergeron J., 1996, A\&A 309, 81

















\end{thebibliography}
\end{document}